\documentclass{amsart}[11pt]
\usepackage[margin=3cm]{geometry}
\usepackage{amsmath,amsfonts, amsthm, amssymb, mathtools}
\usepackage{hyperref}
\usepackage[braket, qm]{qcircuit}
\usepackage{booktabs}
\usepackage{float}

\title[Quantum Computing for Financial Mathematics]{Quantum Computing 
for Financial Mathematics\\
}
\date{\today}

\author{Antoine Jacquier}
\address{Department of Mathematics, Imperial College London, 
and the Alan Turing Institute}
\email{a.jacquier@imperial.ac.uk}

\author{Oleksiy Kondratyev}
\address{Abu Dhabi Investment Authority (ADIA), 
and Department of Mathematics, Imperial College London}
\email{oleksiy.kondratyev@adia.ae}

\author{Gordon Lee}
\email{gtylee@gmail.com}

\author{Mugad Oumgari}
\address{University College London and Lloyds Banking}
\email{Mugad.Oumgari@lloydsbanking.com}


\thanks{AJ is supported by the EPSRC grants EP/W032643/1 and  EP/T032146/1.}

\begin{document}
\maketitle

\begin{abstract}
Quantum computing has recently appeared in the headlines of many scientific 
and popular publications. In the context of quantitative finance, we provide 
here an overview of its potential.
\end{abstract}

\vspace{1cm}

Financial mathematics as a standalone discipline enjoyed many periods of 
rapid development followed by periods of relative calm. As a synthetic 
discipline that exists at the cross-section of applied mathematics, 
financial theory, computer science, prudential regulation, ..., 
it benefits from discoveries and breakthroughs in all these fields.

The birth of financial mathematics is often attributed to the doctoral 
thesis of Louis Bachelier, “The Theory of Speculation”, defended in 1900 
(and written under the supervision of Henri Poincar\'e) and which applied 
a stochastic process (later called a Brownian motion) to the modelling of 
financial assets for the first time. Since then, financial mathematics has 
been influenced by stochastic calculus (It\^o's lemma, Girsanov's theorem), 
control theory (Kalman filter), statistics (Kolmogorov-Smirnov test), 
but also spurred by progresses in microprocessors, financial deregulation, 
ultrafast communication, object-oriented programming, to name just a few.

In this regard, quantum computing opens a new chapter for financial 
mathematics as it promises enormous computing power at very low cost. 
To understand why this is the case, one must appreciate the importance 
of computation in general. At the end of the day, all financial problems 
(pricing, risk management, credit scoring, discovery of trading signals, 
data encryption, portfolio optimisation) are of computational nature, 
and the task of financial mathematics is to provide the most efficient 
and convenient tools to perform this computation.

What is computation? A computation can be defined as a transformation of 
one memory state into another, or a function that transforms information. 
In the case of classical digital computing, the fundamental memory unit is 
a {\em binary digit} (bit) of information. Functions that operate on bits 
of information are called {\em logic gates}, namely Boolean functions that 
can be combined into {\em circuits} capable of performing additions, 
multiplications, and more complex operations. But is the Boolean logic the 
only or even the most general way to realise digital computation? The answer 
is clearly no. Classical computing is just a special case of a more general 
computational framework now called \emph{quantum computing}. A classical bit 
is a two-state system that can exist in either of two discrete deterministic 
states, traditionally denoted~`0' and~`1'. All classical bits are independent 
-- flipping the state of a given bit does not affect the states of other bits. 
These two features of classical computing can be generalised by allowing 
the bit to exist in a superposition of the two states~`0' and~`1', and by 
allowing the states of different bits to be entangled (a certain form of 
correlation). It is also clear how quantum computing obtained its name -- 
superposition and entanglement are the key characteristics of the states 
of quantum systems, and it is tempting to perform computation through the 
{\em controlled evolution of a quantum system}, i.e., by running a 
{\em quantum computer}.

Superposition and entanglement are two features responsible for the 
extraordinary power of quantum computing. They allow for more general 
computation: more general definition of the memory state in comparison 
with classical digital computing and a wider range of possible 
transformations of such memory states. The fundamental memory unit in 
quantum computing is a {\em quantum binary digit} (qubit). Mathematically, 
the state of a qubit is a unit vector in the two-dimensional complex 
vector space. {\em Quantum logic gates} are represented by the 
norm-preserving unitary operators (unitary matrices) acting on qubit 
states. Once the computation is complete (the initial system state has 
been transformed by the {\em quantum circuit} -- a sequence of quantum 
logic gates), the qubit states can be measured (projected onto the 
basis states). Qubits in their basis states correspond to classical 
bits (all superpositions have been collapsed). The rest of the 
computational protocol can be done classically after the read-out of 
the bitstring, which is the output of the quantum computer.

A natural question arises: what is the reason for this superior mode of 
computation not being used until very recently? The answer is that 
although quantum mechanics was formulated almost a century ago (Dirac's 
seminal work "The Principles of Quantum Mechanics" was published in 1930), 
the realisation of the rules of quantum mechanics in the computational 
protocol performed on classical digital computers requires an enormous 
amount of memory. Exponential gains in computing power are offset by 
exponential memory requirements.

In order to perform quantum computations efficiently, we need to use 
actual quantum mechanical systems, with their ability to encode 
information in their states. 
To illustrate this point, the state of a quantum system consisting of~$n$ 
entangled qubits can be described by specifying~$2^n$ probability amplitudes,
a huge amount of information even for very small systems ($n \sim 100$) and 
it would be impossible to store this information in classical memory. 
It took decades of technological progress before Quantum Processing Units 
(QPUs)--–devices that control quantum mechanical systems performing 
computations--–became feasible.

The current state-of-the-art QPUs have several hundred qubits and the qubit 
fidelity is still insufficient for the fault-tolerant computation. However, 
these systems are already large enough and qubit fidelity high enough for 
such quantum computers to be useful. Among many possible quantum computing 
technologies, two qubit types stand out as most developed (and promising):

\begin{itemize}
\item  Qubits made of superconducting circuits 
(coherence time: $\sim 10^3$\textmu$\textrm{s}$)~\cite{somoroff2023quantum}

\begin{table}[H]
\centering
\begin{tabular}{l l | l l}
  \toprule
  one-qubit gate &                 & two-qubit gate &      \\
  \midrule
  Gate time: & $\sim 10^{-2}$\textmu$\textrm{s}$ & 
  Gate time: & $\sim 10^{-2}$-$10^{-1}$\textmu$\textrm{s}$ \\
  Fidelity:  & 99.99\%              & Fidelity:  & 99.9\%   \\
  \bottomrule
\end{tabular}
\end{table}

\item Qubits made of trapped ions (coherence time: 
$>10^8$\textmu$\textrm{s}$)~\cite{bruzewicz2019quantum}

\begin{table}[H]
\centering
\begin{tabular}{l l | l l}
  \toprule
  one-qubit gate &                 & two-qubit gate &    \\
  \midrule
  Gate time: & $\sim 1$-$10$\textmu$\textrm{s}$ & 
  Gate time: & $\sim 10$\textmu$\textrm{s}$              \\
  Fidelity:  & 99.9999\%           & Fidelity:  & 99.9\% \\
  \bottomrule
\end{tabular}
\end{table}

\end{itemize}

These qubit characteristics are indicative of the fact that we approaching
the threshold beyond which various error correction algorithms become
feasible and we may finally enter the era of fault-tolerant quantum computing.

Whilst {\em quantum supremacy} (a term coined by Preskill in 2012) was 
demonstrated by Google in 2019 on a specially designed problem, recent years 
have witnessed clear signs of {\em quantum advantage}---first productive 
applications of quantum computers to real-world problems hard for classical 
computers. 

That said, given the scarcity and cost of full-fledged quantum computers, 
emulators play a pivotal role in the current quantum ecosystem. These emulators 
simulate quantum operations on classical hardware, allowing researchers and 
developers to design, test, and refine quantum algorithms without requiring 
direct access to a quantum machine, moving forward the state of the art in its 
research but by-passing their current limited availability. Graphics Processing 
Units (GPUs), given their parallel processing capabilities, have emerged as 
the go-to hardware for emulating quantum systems. Their architecture is 
well-suited for handling the matrix operations fundamental to quantum mechanics.
Tech giants in recent years have been creating frameworks for quantum computing 
in public: IBM's \texttt{Qiskit} development kit allows any \texttt{python} user 
to implement and test quantum algorithms. Google provides the \texttt{Cirq} 
framework, enabling developers to create, edit, and invoke quantum circuits on 
real and simulated quantum devices. Microsoft's Quantum Development Kit includes 
the~Q$\sharp$ language, which can be used to write quantum algorithms that are 
run on classical simulators. Xanadu's \texttt{pennylane} software framework is 
also specifically designed to implement quantum machine learning tools.

Several areas deserve a particular attention as the most relevant 
financial mathematics problems.\\

\noindent
{\bf Optimisation}\\
Digital quantum computing offers the possibility of solving NP-hard
combinatorial optimisation problems using variational methods such
as the Variational Quantum Eigensolver~\cite{mcclean2016theory, peruzzo2014variational} 
and the Quantum Approximate Optimisation Algorithm~\cite{farhi2014quantum}. 
Both algorithms can be used to solve a wide range of finance-related 
optimisation problems~\cite{jacquier2022quantum}. Moreover, 
variational algorithms are noise-resistant and, therefore,
suitable~\cite{cerezo2021variational, stilck2021limitations} for running 
on the current generation of Noisy Intermediate-Scale Quantum (NISQ) 
computers~\cite{bharti2021noisy, preskill2018quantum}.
Additionally, classically hard optimisation problems naturally lend
themselves to be addressed by {\em analog quantum computers} that 
realise the principles of Adiabatic Quantum Computing. The 
flagship financial use case  is discrete portfolio 
optimisation where we can see the first experimental evidences of
a quantum speedup~\cite{venturelli2019reverse}.\\

\noindent
{\bf Quantum Machine Learning}\\
It is a combination of quantum computing and AI that will likely generate
the most exciting opportunities, including a whole range of possible
applications in finance. We have already witnessed the first promising 
results achieved with parameterised quantum circuits trained as either 
generative models (such as Quantum Circuit Born Machine~\cite{kondratyev2021non}, 
which can be used as a synthetic data generator) or discriminative models 
(such as Quantum Neural Networks~\cite{farhi2018classification} that can be 
trained as classifiers). The possible use cases include market generators, 
data anonymisers, credit scoring, and the generation of trading signals.
The Quantum Generative Adversarial Networks is another type
of generative QML model with significant potential~\cite{assouel2022quantum, chakrabarti2019quantum, huang2021quantum, niu2022entangling, situ2020quantum}.
Similar to classical GANs, a quantum GAN is composed of a generator and of 
a discriminator with the ability to distinguish quantum states. Since each 
quantum state encodes a probability distribution, the quantum GAN 
discriminator can be used to verify whether the datasets in question 
were drawn from the same probability distribution with direct application 
to the time series analysis, detection of structural breaks, and monitoring 
of alpha decay.\\

\noindent
{\bf PDE Solvers}\\
Harrow, Hassidim and Lloyd~\cite{harrow2009quantum} devised a quantum 
algorithm to solve linear systems, beating classical computation 
times. Linear systems are ubiquitous in applications, and many aspects 
of mathematical finance rely on being able to solve such (low- or 
high-dimensional) systems. A particularly important application
is solving Partial Differential Equations (PDEs). A quantum algorithm
for linear PDEs can be used to efficiently price European and Asian 
options in the Black-Scholes framework~\cite{fontanela2021quantum}.\\

\noindent
{\bf Quantum Monte Carlo}\\
Montanaro~\cite{montanaro2015quantum} described a quantum algorithm 
which can accelerate Monte Carlo methods in a very general setting. The
algorithm estimates the expected output value of an arbitrary randomised 
or quantum subroutine with bounded variance, achieving a near-quadratic 
speedup over the best possible classical algorithm. For the recent
advances in quantitative finance applications we refer interested
readers to~\cite{li2023quantum}.\\

\noindent
{\bf Quantum Semidefinite Programming}\\
The key idea behind Quantum Semidefinite Programming (QSDP) is based
on the observation that a normalised positive semidefinite matrix can 
be naturally represented as a quantum state. Operations on quantum 
states can sometimes be computationally cheaper to perform on a 
quantum computer than classical matrix operations. This idea prompted 
the development of quantum algorithms for semidefinite programming.
In finance, QSDP can be used for the maximum risk analysis and 
robust portfolio construction~\cite{jacquier2022quantum}.\\

\vspace{0.5cm}

As a conclusion, after decades of theoretical results and promises, quantum 
computing is now progressively becoming a reality. While full-scale quantum 
computers are not here yet to replace their classical counterparts, they are 
nevertheless already useful both to speed up specific procedures in classical 
algorithms (giving birth to the \emph{hybrid classical-quantum era}) and to 
provide new angles and new ways of thinking about old problems (so-called 
\emph{quantum-inspired algorithms}). Rather than giving way to quantum 
scepticism, we would instead quote David Deustch: \textit{`Quantum 
computation is [$\cdots$] nothing less than a distinctly new way of 
harnessing nature'} [\emph{The Fabric of Reality} (1997), Chapter~9],
and embrace it as a new tool, that will help us apprehend more accurately 
the numerous problems faced in Quantitative Finance. 

\bibliography{References}
\bibliographystyle{siam}

\end{document}